\documentclass{llncs}

\usepackage{paralist}
\usepackage{subcaption}
\usepackage{amsmath}
\usepackage{xfrac}
\usepackage{graphicx}
\usepackage{float}
\usepackage{rotating}
\usepackage{tabularx}
\usepackage{url}
\usepackage{tabulary}
\usepackage{booktabs}
\usepackage{chngcntr}
\usepackage{xspace}
\usepackage{multirow}
\usepackage[title]{appendix}
\usepackage{mdwlist}
\newcommand{\thesystem}{\textsc{Excision}\xspace}

\begin{document}

\title{Include Me Out: In-Browser Detection of Malicious Third-Party Content
Inclusions}

\author{
  Sajjad Arshad, Amin Kharraz, and William Robertson\\
  Northeastern University, Boston, USA\\
  \texttt{\{arshad,mkharraz,wkr\}@ccs.neu.edu}
}

\institute{}

\maketitle

\begin{abstract}

Modern websites include various types of third-party content such as JavaScript,
images, stylesheets, and Flash objects in order to create interactive user
interfaces. In addition to explicit inclusion of third-party content by website
publishers, ISPs and browser extensions are hijacking web browsing sessions with
increasing frequency to inject third-party content (e.g., ads). However,
third-party content can also introduce security risks to users of these websites,
unbeknownst to both website operators and users. Because of the often highly
dynamic nature of these inclusions as well as the use of advanced cloaking
techniques in contemporary malware, it is exceedingly difficult to preemptively
recognize and block inclusions of malicious third-party content before it has
the chance to attack the user's system.

In this paper, we propose a novel approach to achieving the goal of preemptive
blocking of malicious third-party content inclusion through an analysis of
inclusion sequences on the Web. We implemented our approach, called \thesystem,
as a set of modifications to the Chromium browser that protects users from
malicious inclusions while web pages load. Our analysis suggests that by
adopting our in-browser approach, users can avoid a significant portion of
malicious third-party content on the Web. Our evaluation shows that \thesystem
effectively identifies malicious content while introducing a low false positive
rate. Our experiments also demonstrate that our approach does not negatively
impact a user's browsing experience when browsing popular websites drawn from
the Alexa Top 500.

\end{abstract}


\section{Introduction}
\label{sec:intro}

Linking to third-party content has been one of the defining features of the
World Wide Web since its inception, and this feature remains strongly evident
today. For instance, recent research~\cite{ccs2012jsinclusion} reveals that more
than 93\% of the most popular websites include JavaScript from external sources.
Developers typically include third-party content for convenience and performance
-- e.g., many JavaScript libraries are hosted on fast content delivery networks
(CDNs) and are likely to already be cached by users -- or to integrate with
advertising networks, analytics frameworks, and social media. Third-party
content inclusion has also been used by entities other than the website
publishers themselves. For example, ad injection has been adopted by ISPs and
browser extension authors as a prominent technique for monetization~\cite
{adinjection-profit}.

However, the inherent feature of content-sharing on the Web is also an Achilles
heel when it comes to security. Advertising networks, as one example, have
emerged as an important vector for adversaries to distribute attacks to a wide
audience%
~\cite{sp2013linchpins,ccs2012madtracer,www2014shortening,ccs2013spiderweb,imc2014malvertising}.
Moreover, users are more susceptible to \emph{malvertising} in the presence of
ad injection~\cite{usenixsec2015webeval,sp2015adinjection,www2015adinjection}.
In general, linking to third-party content is essentially an assertion of
trust that the content is benign. This assertion can be violated in several
ways, however, due to the dynamic nature of the Web. Since website operators
cannot control external content, they cannot know \emph{a priori} what links
will resolve to in the future. The compromise of linked content or pure
malfeasance on the part of third parties can easily violate these trust
assumptions. This is only exacerbated by the transitive nature of trust on the
Web, where requests for content can be forwarded beyond the first, directly
observable origin to unknown parties.

While the same origin policy (SOP) enforces a modicum of origin-based separation
between code and data from different principals, developers have clamored for
more flexible sharing models provided by, e.g., Content Security Policy
(CSP)~\cite{csp-spec}, Cross-Origin Resource Sharing (CORS)~\cite{cors-spec},
and postMessage-based cross-frame communication. These newer standards permit
greater flexibility in performing cross-origin inclusions, and each come with
associated mechanisms for restricting communication to trusted origins. However,
recent work has shown that these standards are difficult to apply securely in
practice~\cite{ndss2013postman,raid2014csp}, and do not necessarily address the
challenges of trusting remote inclusions on the dynamic Web. In addition to the
inapplicability of some approaches such as CSP, third parties can leverage their
power to bypass these security mechanisms. For example, ISPs and browser
extensions are able to tamper with HTTP traffic to modify or remove CSP rules in
HTTP responses~\cite{usenixsec2015webeval,sp2015adinjection}.

In this paper, we propose an in-browser approach called \thesystem to
automatically detect and block malicious third-party content inclusions as web
pages are loaded into the user's browser or during the execution of browser
extensions. Our approach does not rely on examination of the content of the
resources; rather, it relies on analyzing the sequence of inclusions that leads
to the resolution and loading of a terminal remote resource. Unlike prior
work~\cite{ccs2012madtracer}, \thesystem resolves \emph{inclusion sequences}
through instrumentation of the browser itself, an approach that provides a
high-fidelity view of the third-party inclusion process as well as the ability
to interdict content loading in real-time. This precise view also renders
ineffective common obfuscation techniques used by attackers to evade detection.
Obfuscation causes the detection rate of these approaches to degrade
significantly since obfuscated third-party inclusions cannot be traced using
existing techniques~\cite{ccs2012madtracer}. Furthermore, the in-browser
property of our system allows users to browse websites with a higher confidence
since malicious third-party content is prevented from being included while the
web page is loading.

We implemented \thesystem as a set of modifications to the Chromium browser, and
evaluated its effectiveness by analyzing the Alexa Top 200K over a period of 11
months. Our evaluation demonstrates that \thesystem achieves a 93.39\% detection
rate, a false positive rate of 0.59\%, and low performance overhead. We also
performed a usability test of our research prototype, which shows that
\thesystem does not detract from the user's browsing experience while
automatically protecting the user from the vast majority of malicious content on
the Web. The detection results suggest that \thesystem could be used as a
complementary system to other techniques such as CSP.

The main contributions of this paper are as follows:

\begin{itemize*}

\item We present a novel in-browser approach called \thesystem that automatically
detects and blocks malicious third-party content before it can attack the
user's browser. The approach leverages a high-fidelity in-browser vantage point
that allows it to construct a precise inclusion sequence for every third-party
resource.

\item We describe a prototype of \thesystem for the Chromium browser that can
effectively prevent inclusions of malicious content.

\item We evaluate the effectiveness and performance of our prototype, and show
that it is able to automatically detect and block malicious third-party content
inclusions in the wild -- including malicious resources not previously
identified by popular malware blacklists -- without a significant impact on
browser performance.

\item We evaluate the usability of our prototype and show that most users did
not notice any significant quality impact on their browsing experience.

\end{itemize*}

\section{Problem Statement}
\label{sec:problem-statement}

In the following, we first discuss the threats posed by third-party content and
then motivate our work.

\subsection{Threats}

While the inclusion of third-party content provides convenience for web
developers and allows for integration into advertising distribution, analytics,
and social media networks, it can potentially introduce a set of serious
security threats for users. For instance, advertising networks and social media
have been and continue to be abused as a vector for injection of malware.
Website operators, or publishers, have little control over this content aside
from blind trust or security through isolation. Attacks distributed through
these vectors -- in the absence of isolation -- execute with the same privileges
as all other JavaScript within the security context of the enclosing DOM. In
general, malicious code could launch drive-by downloads~\cite{www2010jsand},
redirect visitors to phishing sites, generate fraudulent clicks on
advertisements~\cite{ccs2012madtracer}, or steal user
information~\cite{ccs2010cssattack}.

Moreover, ad injection has become a new source of income for ISPs and browser
extension authors~\cite{adinjection-profit}. ISPs inject advertisements into web
pages by tampering with their users' HTTP traffic~\cite{isp-adinjection}, and
browser extension authors have recently started to inject or replace ads in web
pages to monetize their work. Ad injection negatively impacts both website
publishers and users by diverting revenue from publishers and exposing users to
malvertising~\cite{sp2015adinjection,www2015adinjection}. In addition to ad
injection, malicious browser extensions can also pose significant risks to users
due to the special privileges they have~\cite{usenixsec2014hulk}.

\subsection{Motivation}

Publishers can try to isolate untrusted third-party content using iframes
(perhaps enhanced with HTML5 sandboxing features), language-based sandboxing, or
policy enforcement%
~\cite{adsafe,ndss2010capabilityleaks,usenixsec2009gatekeeper,usenixsec2010adjail,csf2009langisojs}.
However, these approaches are not commonly used in practice; some degrade the
quality of ads (from the advertiser's perspective), while others are non-trivial
to deploy. Publishers could attempt to use Content Security Policy (CSP)%
~\cite{csp-spec} to define and enforce access control lists for remote
inclusions in the browser. However, due to the dynamic nature of the web, this
approach (and similar access control policy-based techniques) has problems.
Recent studies~\cite{ndss2013postman,raid2014csp} indicate that CSP is difficult
to apply in practice. A major reason for this is the unpredictability of the
origins of inclusions for third-party resources, which complicates the
construction of a correct, yet tight, policy.

For example, when websites integrate third-party advertisements, multiple
origins can be contacted in order to deliver an ad to the user's browser. This
is often due to the practice of re-selling ad space (a process known as ad
syndication) or through real-time ad auctions. Either of these approaches can
result in ads being delivered through a series of JavaScript code
inclusions~\cite{imc2011adexchange}. Additionally, the growing number of browser
extensions makes it a non-trivial task for website operators to enumerate the
set of benign origins from which browser extensions might include a resource. As
an example, for theverge.com website, the number of unique included domains over
a period of 11 months increases roughly linearly; clearly, constructing an
explicit whitelist of domains is a challenging task.

Even if website publishers can keep pace with origin
diversity over time with a comprehensive list of CSP rules, ISPs and browser
extensions are able to tamper with in-transit HTTP traffic and modify CSP rules
sent by the websites. In addition, in browsers such as Chrome, the web page's
CSP does not apply to extension scripts executed in the page's context%
~\cite{csp-content-scripts}; hence, extensions are able to include arbitrary
third-party resources into the web page.

Given the challenges described above, we believe that existing techniques
such as CSP can be evaded and, hence, there is a need for an automatic approach to
protect users from malicious third-party content. We do not necessarily advocate
such an approach in isolation, however. Instead, we envision this approach as a
complementary defense that can be layered with other techniques in order to
improve the safety of the Web.

\section{\thesystem}
\label{sec:architecture}

\begin{figure*}[t]
    \centering
    \includegraphics[width=1\textwidth]{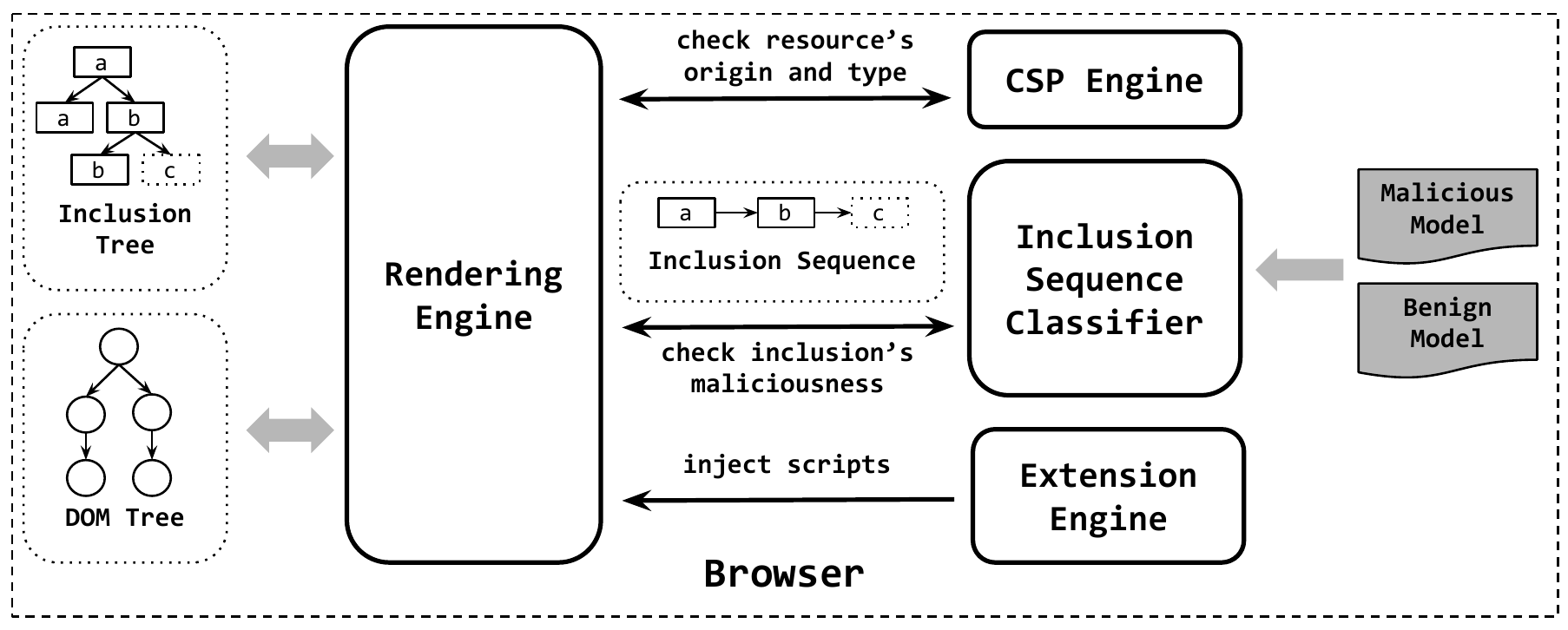}
    \caption{An overview of \thesystem.}
    \label{fig:architecture}
\end{figure*}

In this section, we describe \thesystem, our approach for detecting and blocking
the inclusion of malicious third-party content in real-time. An overview of our
system is shown in Figure~\ref{fig:architecture}. \thesystem operates by
extracting resource \emph{inclusion trees} from within the browser. The
inclusion tree precisely records the inclusion relationships between different
resources in a web page. When the user requests a web page, the browser
retrieves the corresponding HTML document and passes it to the rendering engine.
The rendering engine incrementally constructs an inclusion tree for the DOM and
begins extracting external resources such as scripts and frames as it reaches
new HTML tags. For inclusion of a new resource, the rendering engine consults
the CSP engine and the \emph{inclusion sequence classifier} in order to decide
whether to include the resource. If the resource's origin and type are
whitelisted in the CSP rules, the rendering engine includes the resource without
consulting the inclusion sequence classifier and continues parsing the rest of
the HTML document. Otherwise, it extracts the \emph{inclusion sequence} (path
through the page's inclusion tree) for the resource and forwards this to the
inclusion sequence classifier. Using pre-learned models, the classifier returns
a decision about the malice of the resource to the rendering engine. Finally,
the rendering engine discards the resource if it was identified as malicious.
The same process occurs for resources that are included dynamically during the
execution of extension scripts after they are injected into the page.

\subsection{Inclusion Trees and Sequences}
\label{sec:inclusion_tree}

\begin{figure}[t]
   \centering
   \includegraphics[width=1\textwidth]{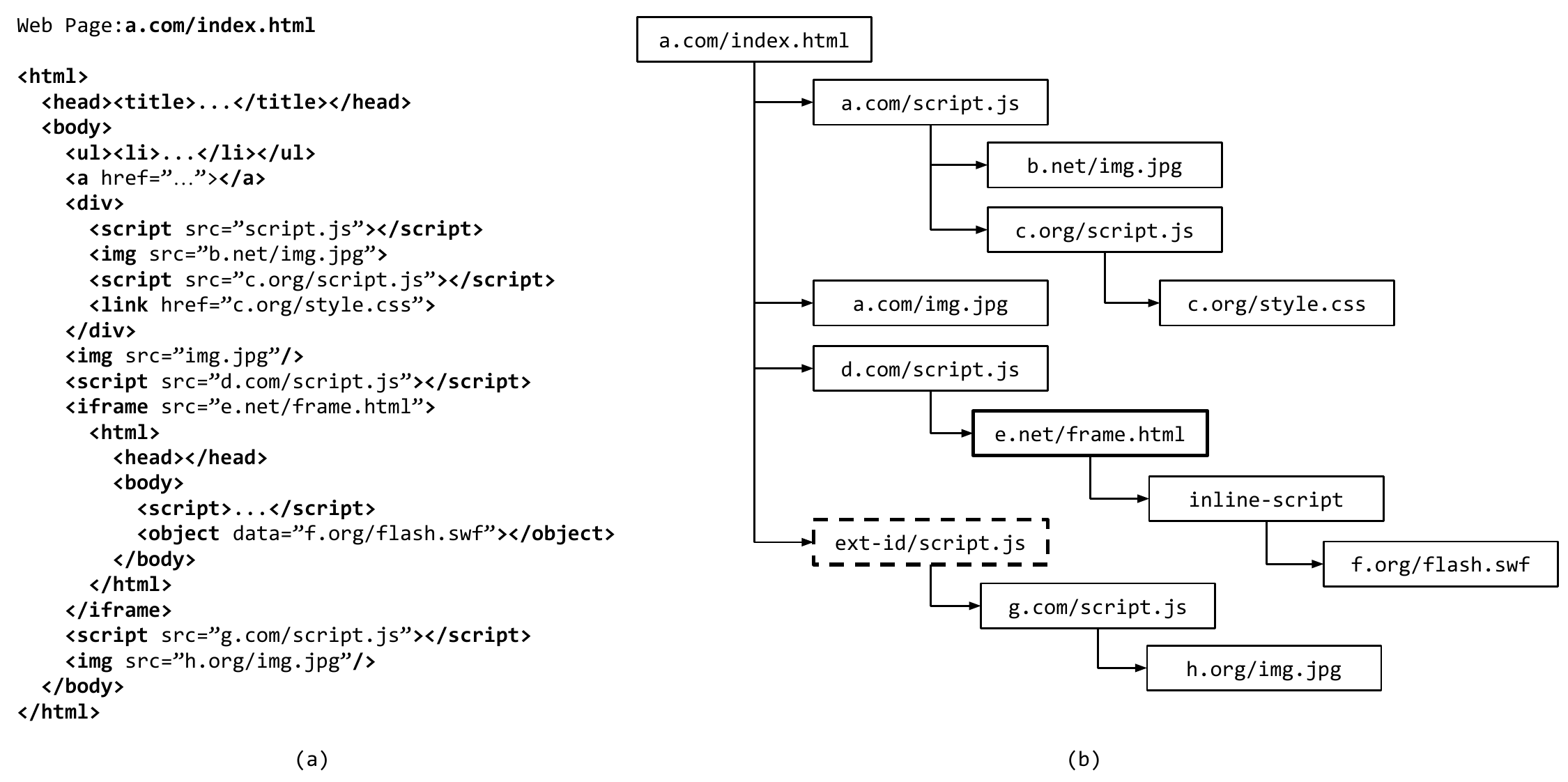}
   \caption{(a) DOM Tree, and (b) Inclusion Tree}
   \label{fig:dom-inclusion-tree}
\end{figure}

A website can include resources in an HTML document from any origin so long as
the inclusion respects the same origin policy, its standard exceptions, or any
additional policies due to the use of CSP, CORS, or other access control
framework. A first approximation to understanding the inclusions of third-party
content for a given web page is to process its DOM tree~\cite{domtree} while the
page loads. However, direct use of a web page's DOM tree is unsatisfactory because
the DOM does not in fact reliably record the inclusion relationships between
resources referenced by a page. This follows from the ability for JavaScript to
manipulate the DOM at run-time using the DOM API.

Instead, in this work we define an \emph{inclusion tree} abstraction extracted
directly from the browser's resource loading code. Unlike a DOM tree, the
inclusion tree represents how different resources are included in a web page
that is invariant with respect to run-time DOM updates. It also discards
irrelevant portions of the DOM tree that do not reference remote content. For
each resource in the inclusion tree, there is an \emph{inclusion sequence} that
begins with the root resource (i.e., the URL of the web page) and terminates
with the corresponding resource. Furthermore, browser extensions can also
manipulate the web page by injecting and executing JavaScript code in the page's
context. Hence, the injected JavaScript is considered a direct child of the root
node in the inclusion tree. An example of a DOM tree and its corresponding
inclusion tree is shown in Figure~\ref{fig:dom-inclusion-tree}. As shown in
Figure~\ref{fig:dom-inclusion-tree}b, \texttt{f.org/flash.swf} has been
dynamically added by an \texttt{inline script} to the DOM tree, and its
corresponding inclusion sequence has a length of 4 since we remove the inline
resources from inclusion sequence. Moreover, \texttt{ext-id/script.js} is
injected by an extension as the direct child of the root resource. This script
then included \texttt{g.com/script.js}, which in turn included
\texttt{h.org/img.jpg}.

\subsection{Inclusion Sequence Classification}
\label{sec:classification}

Given an inclusion sequence, \thesystem must classify it as benign or malicious
based on features extracted from the sequence. The task of the \emph{inclusion
sequence classifier} is to assign a class label from the set
\textsf{\{benign,malicious\}} to a given sequence based on previously learned
models from a labeled data set. In our definition, a malicious sequence is one
that starts from the root URL of a web page and terminates in a URL that
delivers malicious content. For classification, we used hidden Markov models
(HMM)~\cite{ieee1989hmm}. Models are comprised of states, each of which holds
transitions to other states based on a probability distribution. Each state can
probabilistically emit a symbol from an alphabet. There are other sequence
classification techniques such as Na\"{i}ve Bayes~\cite{ecml1998naivebayes}, but
we used an HMM for our classifier because we also want to model the
inter-dependencies between the resources that compose an inclusion sequence.

In the training phase, the system learns two HMMs from a training set of labeled
sequences, one for the benign class and one for the malicious class. We
estimated the HMM parameters by employing the Baum-Welch algorithm which finds
the maximum likelihood estimate of these parameters based on the set of observed
sequences. In our system, we empirically selected 20 for the number of states
that are fully connected to each other. In the subsequent detection phase, we
compute the likelihood of a new sequence given the trained models using the
forward-backward algorithm and assign the sequence to the class with the highest
likelihood. Training hidden Markov models is computationally expensive. However,
computing the likelihood of a sequence is instead very efficient, which makes it
a suitable method for real-time classification~\cite{ieee1989hmm}.

\section{Classification Features}
\label{sec:features}

Let $r_0 \rightarrow r_1 \rightarrow \dots \rightarrow r_n$ be an inclusion
sequence as described above. Feature extraction begins by converting the
inclusion sequence into sequences of feature vectors. After analyzing the
inclusion trees of several thousand benign and malicious websites for a period
of 11 months, we identified 12 feature types from three categories. For each
feature type, we compute two different features: individual and relative
features. An individual feature value is only dependent on the current resource,
but a relative feature value is dependent on the current resource and its
preceding (or parent) resources. Consequently, we have 24 features for each
resource in an inclusion sequence. Individual features can have categorical or
continuous values. All continuous feature values are normalized on
$\left[0,1\right]$ and their values are discretized. In the case of continuous
individual features, the relative feature values are computed by comparing the
individual value of the resource to its parent's individual value. The result of
the comparison is \textsf{less}, \textsf{equal}, or \textsf{more}. We use the
value \textsf{none} for the root resource. To capture the high-level
relationships between different inclusions, we only consider the host part of
the URL for feature calculation.

\subsection{DNS-based Features}

The first feature category that we consider is based on DNS properties of the
resource host.

\begin{table}[t]
    \centering
    \begin{minipage}{.41\textwidth}
		\setlength{\tabcolsep}{10pt}
		\fontsize{8}{8}\selectfont
	    \begin{tabular*}{\linewidth}{@{}ll@{}}
	    \toprule
	    \textbf{Value} & \textbf{Example} \\
	    \midrule
	    \texttt{none} & IPs, Extensions \\
	    \texttt{gen} & *.com, *.org \\
	    \texttt{gen-subdomain} & *.us.com \\
	    \texttt{cc} & *.us, *.de, *.cn \\
	    \texttt{cc-subdomain} & *.co.uk, *.com.cn \\
	    \texttt{cc-int} & *.xn{-{}-}p1ai (ru) \\
	    \texttt{other} & *.biz, *.info \\
	    \bottomrule
	    \end{tabular*}
    \caption{Individual TLD values}
    \label{tab:tld:individual}
	\end{minipage}\hfill
	\begin{minipage}{.55\textwidth}
		\fontsize{8}{8}\selectfont
		\setlength{\tabcolsep}{10pt}
	    \begin{tabular*}{\linewidth}{@{}ll@{}}
	    \toprule
	    \textbf{Value} & \textbf{Example} \\
	    \midrule
	    \texttt{none} & root resource \\
	    \texttt{\{got,lost\}-tld} & Ext. $\rightarrow$ *.de, *.us $\rightarrow$ IP \\
	    \texttt{gen-to-\{cc,other\}} & *.org $\rightarrow$ \{*.de, *.info\} \\
	    \texttt{cc-to-\{gen,other\}} & *.uk $\rightarrow$ \{*.com, *.biz\} \\
	    \texttt{other-to-\{gen,cc\}} & *.info $\rightarrow$ \{*.net, *.uk\} \\
	    \texttt{same-\{gen,cc,other\}} & *.com $\rightarrow$ *.com \\
	    \texttt{diff-\{gen,cc,other\}} & *.info $\rightarrow$ *.biz \\
	    \bottomrule
	    \end{tabular*}
    \caption{Relative TLD values.}
    \label{tab:tld:relative}
\end{minipage}\hfill
\end{table}

\begin{table}[t]
    \centering
    \begin{minipage}{.41\textwidth}
		\setlength{\tabcolsep}{8pt}
		\fontsize{8}{8}\selectfont
	    \begin{tabular*}{\linewidth}{@{}ll@{}}
	    \toprule
	    \textbf{Value} & \textbf{Example} \\
	    \midrule
	    \texttt{ipv6} & 2607:f0d0::::4 \\
	    \texttt{ipv4-private} & 192.168.0.1 \\
	    \texttt{ipv4-public} & 4.2.2.4 \\
	    \texttt{extension} & Ext. Scripts \\
	    \texttt{dns-sld} & google.com \\
	    \texttt{dns-sld-sub} & www.google.com \\
	    \texttt{dns-non-sld} & abc.dyndns.org \\
	    \texttt{dns-non-sld-sub} & a.b.dyndns.org \\
	    \bottomrule
	    \end{tabular*}
    \caption{Individual type values}
    \label{tab:type:individual}
	\end{minipage}\hfill
	\begin{minipage}{.56\textwidth}
		\setlength{\tabcolsep}{10pt}
		\fontsize{8}{8}\selectfont
	    \begin{tabular*}{\linewidth}{@{}ll@{}}
	    \toprule
	    \textbf{Value} & \textbf{Example} \\
	    \midrule
	    \texttt{none} & root resource \\
	    \texttt{same-site} & w.google.com $\rightarrow$ ad.google.com \\
	    \texttt{same-sld} & 1.dyndns.org $\rightarrow$ 2.dyndns.org \\
	    \texttt{same-company} & ad.google.com $\rightarrow$ www.google.de \\
	    \texttt{same-eff-tld} & bbc.co.uk $\rightarrow$ london.co.uk \\
	    \texttt{same-tld} & bbc.co.uk $\rightarrow$ london.uk \\
	    \texttt{different} & google.com $\rightarrow$ facebook.net \\
	    \\
	    \bottomrule
	    \end{tabular*}
    \caption{Relative type values.}
    \label{tab:type:relative}
\end{minipage}\hfill
\end{table}

\textbf{Top-Level Domain.}
For this feature, we measure the types of TLDs from which a resource is included
and how it changes along the inclusion sequence. For every resource in an
inclusion sequence, we assign one of the values in
Table~\ref{tab:tld:individual} as an individual feature. For the relative
feature, we consider the changes that occur between the top-level domain of the
preceding resource and the resource itself. Table~\ref{tab:tld:relative} shows
15 different values of the relative TLD feature.

\textbf{Type.}
This feature identifies the types of resource hosts and their changes along
the inclusion sequence. Possible values of individual and relative features are
shown in Table~\ref{tab:type:individual} and Table~\ref{tab:type:relative}
respectively.

\textbf{Level.}
A domain name consists of a set of labels separated by dots. We say a domain
name with $n$ labels is in level $n-1$. For example, \texttt{www.google.com} is
in level 2. For IP addresses and extension scripts, we consider their level to
be 1. For a given host, we compute the individual feature by dividing the level
by a maximum value of 126.

\textbf{Alexa Ranking.}
We also consider the ranking of a resource's domain in the Alexa Top 1M
websites. To compute the normalized ranking as an individual feature, we divide
the ranking of the domain by one million. For IP addresses, extensions, and
domains that are not in the top 1M, we use the value \textsf{none}.

\subsection{String-based Features}

We observed that malicious domain names often make liberal use of digits and
hyphens in combination with alphabetical characters. So, in this feature
category, we characterize the string properties of resource hosts. For IP
addresses and extension scripts, we assign the value 1 for individual features.

\textbf{Non-Alphabetic Characters.}
For this feature, we compute the individual feature value by dividing the number
of non-alphabetical characters over the length of domain.

\textbf{Unique Characters.}
We also measure the number of unique characters that are used in a domain. The
individual feature is the number of unique characters in the domain divided by
the maximum number of unique characters in the domain name, which is 38 (26
alphabetics, 10 digits, hyphen, and dot).

\textbf{Character Frequency.}
For this feature, we simply measure how often a single character is seen in a
domain. To compute an individual feature value, we calculate the frequency of
each character in the domain and then divide the average of these frequencies by
the length of the domain to normalize the value.

\textbf{Length.}
In this feature, we measure the length of the domain divided by the maximum
length of a domain, which is 253.

\textbf{Entropy.}
In practice, benign domains are typically intended to be
memorable to users. This is often not a concern for attackers, as evidenced by
the use of domain generation algorithms~\cite{ndss2011exposure}. Consequently,
we employ Shannon entropy to measure the randomness of domains in the inclusion
sequence. We calculate normalized entropy as the absolute Shannon entropy
divided by the maximum entropy for the domain name.

\subsection{Role-based Features}

We observed that identifying the role of resources in the inclusion sequences
can be helpful in detecting malicious resources. For example, recent
work~\cite{www2014shortening} reveals that attackers misuse ad networks as well
as URL shortening services for malicious intent. So far, we consider three
roles for a resource:%
\begin{inparaenum}[\itshape i)\upshape]
     \item ad-network,
     \item content delivery network (CDN), and
     \item URL shortening service.
\end{inparaenum}%

In total, we have three features in this category, as each host can
simultaneously perform multiple roles. Both individual and relative features in
this category have binary values. For the individual feature, the value is
\textsf{Yes} if the host has the role, and \textsf{No} otherwise. For the
relative feature, we assign a value \textsf{Yes} if at least one of the
preceding hosts have the corresponding role, and \textsf{No} otherwise. For
extension scripts, we assign the value \textsf{No} for all of the features. To
assign the roles, we compiled a list of common hosts related to these roles that
contains 5,767 ad-networks, 48 CDNs, and 461 URL shortening services.

\section{Implementation}
\label{sec:implementation}

In this section, we discuss our prototype implementation of \thesystem for
detecting and blocking malicious third-party content inclusions. We implemented
\thesystem as a set of modifications to the Chromium browser\footnote{While our
implementation could be adopted as-is by any browser vendors that use
WebKit-derived engines, the design presented here is highly likely to be
portable to other browsers.}. In order to implement our system, we needed to
modify Blink and the Chromium extension engine to enable \thesystem to detect
and block inclusions of malicious content in an online and automatic fashion
while the web page is loading. The entire set of modifications consists of less
than 1,000 lines of C++ and several lines of JavaScript.

\subsection{Enhancements to Blink}

Blink is primarily responsible for parsing HTML documents, managing script
execution, and fetching resources from the network. Consequently, it is ideally
suited for constructing the inclusion tree for a web page, as well as blocking
the inclusion of malicious content.

\textbf{Tracking Resource Inclusion.}
Static resource inclusions that are hard-coded by publishers inside the page's
HTML are added to the inclusion tree as the direct children of the root node.
For dynamic inclusions (e.g.,~via the \textsf{createElement()} and
\textsf{write()} DOM API functions), the system must find the script resource
responsible for the resource inclusion. To monitor dynamic resource inclusions,
the system tracks the start and termination of script execution. Any resources
that are included in this interval will be considered as the children of that
script resource in the inclusion tree.

\textbf{Handling Events and Timers.}
Events and timers are widely used by web developers to respond to user
interactions (e.g.,~clicking on an element) or schedule execution of code after
some time has elapsed. To capture the creation and firing of events and timers,
the system tracks the registration of callback functions for the corresponding
APIs.

\subsection{Enhancements to the Chromium Extension Engine}

The Chromium extension engine handles the loading, management, and execution of
extensions. To access the page's DOM, the extension injects and executes
\emph{content scripts} in the page's context which are regular JavaScript
programs.

\textbf{Tracking Content Scripts Injection and Execution.}
Content scripts are usually injected into web pages either via the extension's
manifest file using the \textsf{content\_scripts} field or at runtime via the
\textsf{executeScript} API. Either way, content scripts are considered direct
children of the root node in the inclusion tree. Therefore, in order to track
the inclusion of resources as a result of content script execution, the
extension engine was modified to track the injection and execution of content
scripts.

\textbf{Handling Callback Functions.}
Like any other JavaScript program, content scripts rely heavily on callback
functions. For instance, \textsf{onMessage} and \textsf{sendMessage} are used by
content scripts to exchange messages with their background pages. To track the
execution of callback functions, two JavaScript files were modified in the
extension engine which are responsible for invocation and management of callback
functions.

\section{Evaluation}
\label{sec:evaluation}

In this section, we evaluate the security benefits, performance, and usability
of the \thesystem prototype. We describe the data sets we used to train and
evaluate the system, and then present the results of the experiments.

\subsection{Data Collection}

\begin{table}[t]
    \centering
    \setlength{\tabcolsep}{3pt}
    {
    \fontsize{8}{8}\selectfont
    \tabcolsep 27.5pt
    \begin{tabular*}{\linewidth}{@{}lrr@{}}
    \toprule
    \textbf{Item} & \textbf{Website Crawl} & \textbf{Extension Crawl} \\
    \midrule
    Websites Crawled & 234,529 & 20 \\
    Unavailable Websites & 7,412 & 0 \\
    \midrule
    Unique Inclusion Trees & 47,789,268 & 35,004 \\
    Unique Inclusion Sequences & 27,261,945 & 61,489 \\
    \midrule
    Unique URLs & 546,649,590 & 72,064 \\
    Unique Hosts & 1,368,021 & 1,144 \\
    Unique Sites & 459,615 & 749 \\
    Unique SLDs & 419,119 & 723 \\
    Unique Companies & 384,820 & 719 \\
    Unique Effective TLDs & 1,115 & 21 \\
    Unique TLDs & 404 & 21 \\
    Unique IPs & 9,755 & 3 \\
    \bottomrule
    \end{tabular*}}
    \caption{Summary of crawling statistics.}
    \label{tab:crawling-statistics}
\end{table}

\begin{table}[t]
    \centering
    \setlength{\tabcolsep}{14.5pt}
    {
    \fontsize{8}{8}\selectfont
    \begin{tabular*}{\textwidth}{@{}lcccc@{}}
    \toprule
    \multirow{2}{*}{\textbf{Dataset}} & \multicolumn{2}{c}{\textbf{No. of Inclusion Sequences}} & \multicolumn{2}{c}{\textbf{No. of Terminal hosts}} \\
    \cmidrule[0.5pt](ll){2-5}
    & \textbf{Web. Crawl} & \textbf{Ext. Crawl} & \textbf{Web. Crawl} & \textbf{Ext. Crawl} \\
    \midrule
    Benign & 3,706,451 & 7,372 & 35,044 & 250 \\
    Malicious & 25,153 & 19 & 1,226 & 2 \\
    \bottomrule
    \end{tabular*}}
    \caption{Data sets used in the evaluation.}
    \label{tab:dataset-statistics}
\end{table}

To collect inclusion sequences, we performed two separate crawls for websites
and extensions. The summary of crawling statistics are presented in
Table~\ref{tab:crawling-statistics}.

\textbf{Website Crawl.}
We built a crawler based on an instrumented version of
PhantomJS~\cite{phantomjs}, a scriptable open source browser based on WebKit,
and crawled the home pages of the Alexa Top 200K. We performed our data
collection from June 20th, 2014 to May 11th, 2015. The crawl was parallelized by
deploying 50 crawler instances on five virtual machines, each of which crawled a
fixed subset of the Alexa Top 200K websites. To ensure that visited websites did
not store any data on the clients, the crawler ran a fresh instance of PhantomJS
for each visit. Once all crawlers finished crawling the list of websites, the
process was restarted from the beginning. To thwart cloaking
techniques~\cite{sec2011deseo} utilized by attackers, the crawlers presented a
user agent for IE 6.0 on Windows and employed Tor to send HTTP requests from
different source IP addresses. We also address JavaScript-based browser
fingerprinting by modifying the internal implementation of the
\texttt{navigator} object to return a fake value for the \texttt{appCodeName},
\texttt{appName}, \texttt{appVersion}, \texttt{platform}, \texttt{product},
\texttt{userAgent}, and \texttt{vendor} attributes.

\textbf{Extension Crawl.}
To collect inclusion sequences related to extensions, we used 292 Chrome
extensions reported in prior work~\cite{www2015adinjection} that injected ads
into web pages. Since ad-injecting extensions mostly target shopping websites
(e.g., Amazon), we chose the Alexa Top 20 shopping websites for crawling to
trigger ad injection by those 292 extensions. We built a crawler by
instrumenting Chromium~43 and collected data for a period of one week from June
16th to June 22nd, 2015. The system loaded every extension and then visited the
home pages of the Alexa Top 20 shopping websites using Selenium
WebDriver~\cite{selenium}. This process was repeated after crawling the entire
set of extensions. In addition, our crawler triggered all the events and timers
registered by content scripts.

\subsection{Building Labeled Datasets}
\label{sec:building-labelled-dataset}

To classify a given inclusion sequence as benign or malicious, we trained two
hidden Markov models for benign and malicious inclusion sequences from our data
set. We labeled collected inclusion sequences as either benign or malicious
using VirusTotal~\cite{virustotal}. VirusTotal's URL scanning service aggregates
reports of malicious URLs from most prominent URL scanners such as Google Safe
Browsing~\cite{gsb} and the Malware Domain List. The malicious data set contains
all inclusion sequences where the last included resource's host is reported
malicious by at least three out of the 62 URL scanners in VirusTotal. On the
other hand, the benign data set only contains inclusion sequences that do not
contain any host in the entire sequence that is reported as malicious by any URL
scanner in VirusTotal. To build benign data set, we considered reputable domains
such as well-known search engines and advertising networks as benign regardless
of whether they are reported as malicious by any URL scanner in VirusTotal.
Table~\ref{tab:dataset-statistics} summarizes the data sets. The unique number
of inclusion sequences and terminal hosts are shown separately for the website
and extension data sets. The terminal hosts column is the number of unique hosts
that terminate inclusion sequences.

\subsection{Detection Results}
\label{sec:detection_results}

To evaluate the accuracy of our classifier, we used 10-fold cross-validation, in
which we first partitioned each data set into 10 equal-sized folds, trained the
models on nine folds, and then validated the resulting models with the remaining
fold. The process was repeated for each fold and, at the end, we calculated the
average false positive rate and false negative rate. When splitting the data set
into training and testing sets, we made sure that inclusion sequences with
different lengths were present in both. We also ensured that both sets contained
extension-related inclusion sequences.

The results show that our classifier achieved a false positive rate of 0.59\%
and false negative rate of 6.61\% (detection rate of 93.39\%). Most of the false
positives are due to inclusion sequences that do not appear too often in the
training sets. Hence, users are unlikely to experience many false positives in a
real browsing environment (as will be shown in our usability analysis in
Section~\ref{sec:usability}).

\begin{figure}[t]
	\centering
	\begin{subfigure}[b]{0.49\textwidth}
		\includegraphics[width=\linewidth]{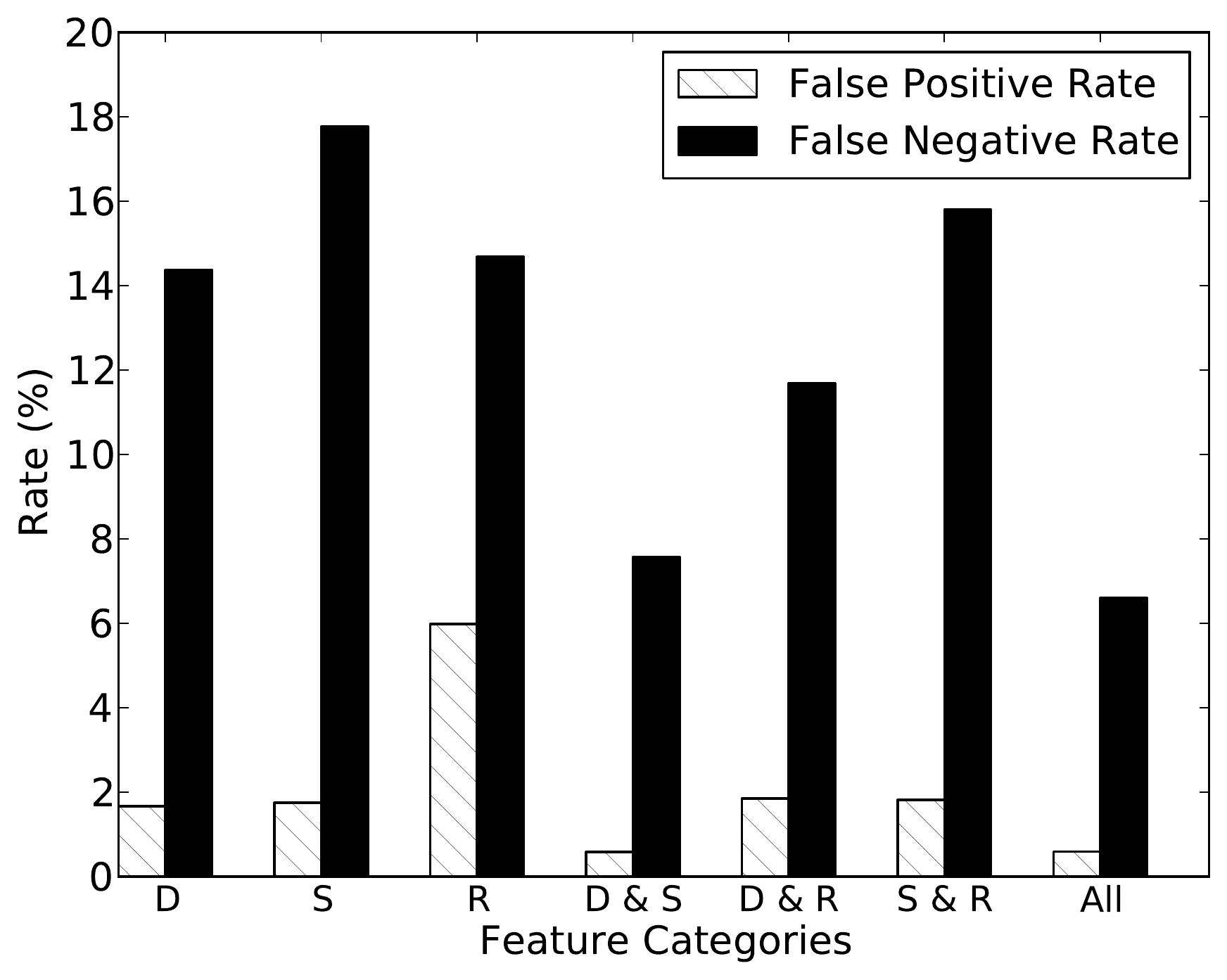}
		\caption{Effectiveness of features for classification (D = DNS, S =
		String, R = Role).}
		\label{fig:eval:detection-rate}
	\end{subfigure}
	\begin{subfigure}[b]{0.49\textwidth}
	\centering
		\includegraphics[width=\linewidth]{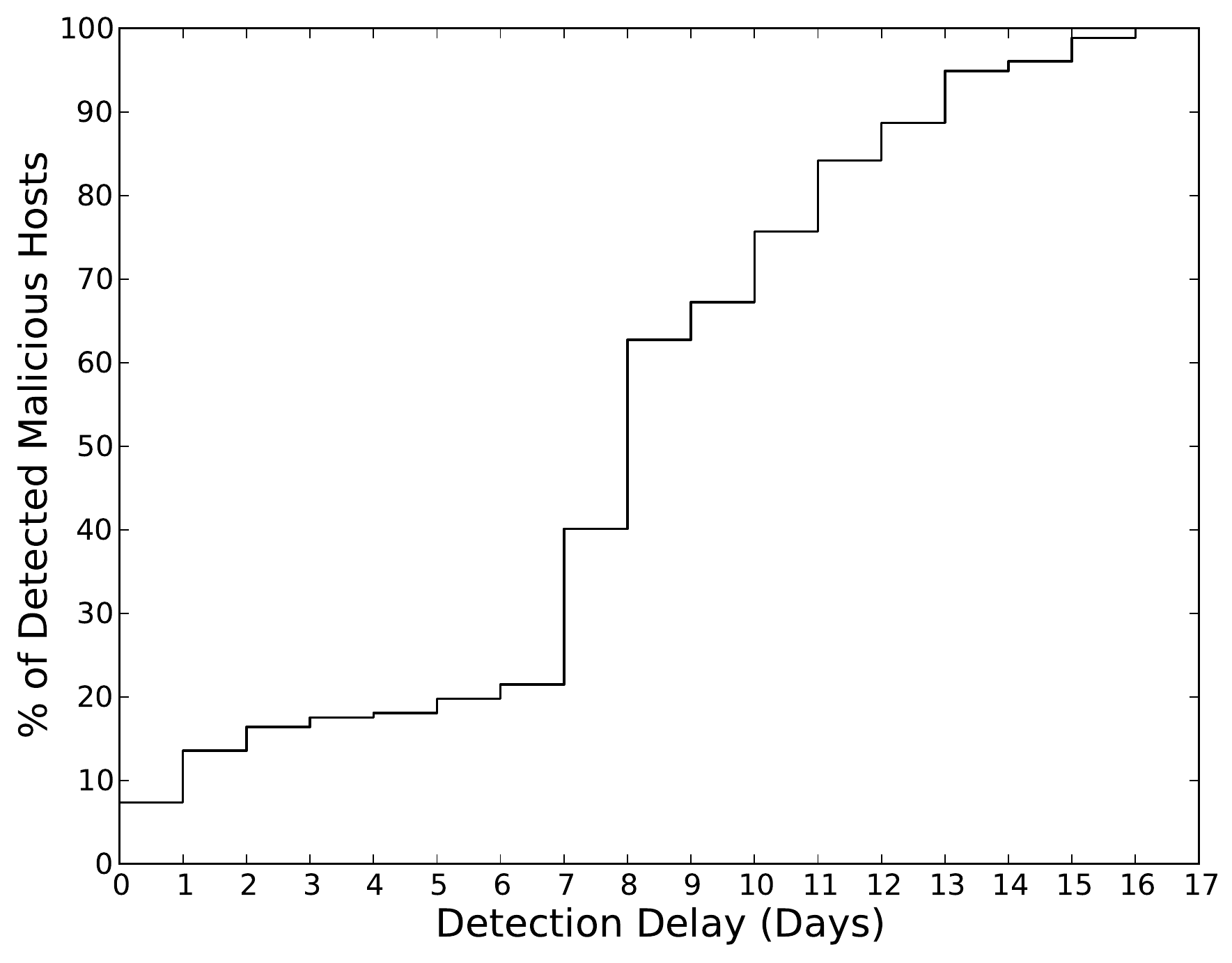}
		\caption{Early detection results.\newline}
		\label{fig:eval:early-detection}
	\end{subfigure}
	\caption{Feature category contributions and early detection results.}
	\label{fig:eval}
\end{figure}

To quantify the contribution of different feature categories to the
classification, we trained classifiers using different combinations of feature
categories and compared the results. Figure~\ref{fig:eval:detection-rate} shows
the false positive rate and false negative rate of every combination with a
10-fold cross-validation training scheme. According to Figure%
~\ref{fig:eval:detection-rate}, the best false positive and false negative rates
were obtained using a combination of all feature categories.

\subsection{Comparison with URL Scanners} \label{sec:comparison}

To evaluate the ability of our system in detecting unreported suspicious
hosts, we ran our classifier on inclusion sequences collected from June 1st
until July 14th, 2015. We compared our detection results with reports from URL
scanners in VirusTotal and detected 89 new suspicious hosts. We believe that
these hosts are in fact dedicated malicious hosts that play the role of
redirectors and manage malicious traffic flows as described in prior
work~\cite{sp2013linchpins}. These hosts did not deliver malicious resources
themselves, but they consistently included resources from other hosts that were
flagged as malicious by URL scanners. Out of 89 suspicious domains, nearly 44\%
were recently registered in 2015, and more than 23\% no longer resolve to an IP
address.

Furthermore, we detected 177 hosts that were later reported by URL scanners
after some delay. Figure~\ref{fig:eval:early-detection} shows the early
detection results of our system. A significant number of these hosts were not
reported until some time had passed after \thesystem initially identified them.
For instance, nearly 78\% of the malicious hosts were not reported by any URL
scanner during the first week.

\subsection{Performance}
\label{sec:performance}

To assess the performance of \thesystem, we used Selenium WebDriver to
automatically visit the Alexa Top 1K with both original and modified Chromium
browsers. In order to measure our prototype performance with a realistic set of
extensions, we installed five of the most popular extensions in the Chrome Web
Store: Adblock Plus, Google Translate, Google Dictionary, Evernote Web Clipper,
and Tampermonkey.

For each browser, we visited the home pages of the entire list of websites and
recorded the total elapsed time. Due to the dynamic nature of ads and their
influence on page load time, we repeated the experiment 10 times and measured
the average elapsed time. On average, the elapsed times were 3,065 and 3,438
seconds for the original and modified browsers respectively. Therefore,
\thesystem incurred a 12.2\% overhead on browsing time on average, which
corresponds to a noticeable overhead that is nevertheless acceptable for many
users (see Section~\ref{sec:usability}). To measure the overhead incurred by
\thesystem on browser startup time, we launched the modified browser 10 times
and measured the average browser launch time. \thesystem caused a 3.2 seconds
delay on browser startup time, which is ameliorated by the fact that this is a
one-time performance hit.

\subsection{Usability}
\label{sec:usability}

We conducted an experiment to evaluate the impact of \thesystem on the user's
browsing experience. We conducted the study on 10 students that self-reported as
expert Internet users. We provided each participant with a list of 50 websites
that were selected randomly from the Alexa Top 500 and then asked them to visit
at least three levels down in each website. Participants were asked to report
the number of visited pages and the list of domains reported as malicious by our
system. In addition, participants were asked to record the number of errors they
encountered while they browsed the websites. Errors were considered to occur
when the browser crashed, the appearance of a web page was corrupted, or page
load times were abnormally long. Furthermore, in order to ensure that benign
extensions were not prevented from executing as expected in the presence of our
system, the browser was configured to load the five popular extensions listed in
Section~\ref{sec:performance} and participants were asked to report any problem
while using the extensions.

The results of the study show that out of 5,129 web pages visited by the
participants, only 83 errors were encountered and the majority of web pages
loaded correctly. Most of these errors happened due to relatively high load
times. In addition, none of the participants reported any broken extensions.
Furthermore, 31 malicious inclusions were reported by our tool that were
automatically processed (without manual examination, for privacy reasons) using
VirusTotal. Based on the results, we believe that our proof-of-concept prototype
is compatible with frequently used websites and extensions, and can be improved
through further engineering to work completely free of errors.

\textbf{Ethics.}
In designing the usability experiment, we made a conscious effort to avoid
collecting personal or sensitive information. In particular, we restricted the
kinds of information we asked users to report to incidence counts for each of
the categories of information, except for malicious URLs that were reported by
our tool. Malicious URLs were automatically submitted to VirusTotal to obtain a
malice classification before being discarded, and were not viewed by us or
manually inspected. In addition, the participants were asked to avoid browsing
websites requiring a login or involving sensitive subject matter.

\section{Related Work}
\label{sec:related-work}

\textbf{Third-party Content Isolation.}
Several recent research
projects~\cite{sp2008op,usenixosdi2010illinois,usenixsec2009gazelle} attempted
to improve the security of browsers by isolating browser components in order to
minimize data sharing among software components. The main issue with these
approaches is that they do not perform any isolation between JavaScript loaded
from different domains and web applications, letting untrusted scripts access
the main web application's code and data. Efforts such as
AdJail~\cite{usenixsec2010adjail} attempt to protect privacy by isolating ads
into an iframe-based sandbox. However, this approach restricts contextual
targeting advertisement in which ad scripts need to have access to host page
content.

\textbf{Detecting Malicious Domains.}
There are multiple approaches to automatically detecting malicious web domains.
Madtracer~\cite{ccs2012madtracer} has been proposed to automatically capture
malvertising cases. But, this system is not as precise as our approach in
identifying the causal relationships among different domains.
EXPOSURE~\cite{ndss2011exposure} employs passive DNS analysis techniques to
detect malicious domains. SpiderWeb~\cite{ccs2013spiderweb} is also a system
that is able to detect malicious web pages by crowd-sourcing redirection chains.
Segugio~\cite{dsn2015segugio} tracks new malware-control domain names in very
large ISP networks. WebWitness~\cite{usenixsec2015webwitness} automatically
traces back malware download paths to understand attack trends. While these
techniques can be used to automatically detect malicious websites and update
blacklists, they are not online systems and may not be effectively used to
detect malicious third-party inclusions since users expect a certain level of
performance while browsing the Web.

Another effective detection approach is to produce blacklists of malicious sites
by scanning the Internet that can be efficiently checked by the browser (e.g.,
Google Safe Browsing~\cite{gsb}). Blacklist construction requires extensive
infrastructure to continuously scan the Internet and bypass cloaking and general
malware evasion attempts in order to reliably identify malware distribution
sites, phishing pages, and other Web malice. As our evaluation in
Section~\ref{sec:evaluation} demonstrates, these blacklists sometimes lag the
introduction of malicious sites on the Internet, or fail to find these malicious
sites. However, they are nevertheless effective, and we view the approach we
propose as a complementary technique to established blacklist generation and
enforcement techniques.

\textbf{Policy Enforcement.}
Another approach is to search and restrict third-party code included in web
applications~%
\cite{ndss2010capabilityleaks,usenixsec2009gatekeeper,csf2009langisojs}. For
example, ADsafe~\cite{adsafe} removes dangerous JavaScript features (e.g.,
\texttt{eval}), enforcing a whitelist of allowed JavaScript functionality
considered safe. It is also possible to protect against malicious JavaScript ads
by enforcing policies at
runtime~\cite{asiaccs2009lightjs,usenixosdi2006browsershield}. For example,
Meyerovich et al.~\cite{sp2010conscript} introduce a client-side framework that
allows web applications to enforce fine-grained security policies for DOM
elements. AdSentry~\cite{acsac2011adsentry} provides a shadow JavaScript engine
that runs untrusted ad scripts in a sandboxed environment.

\section{Conclusion}
\label{sec:conclusion}

In this paper, we presented \thesystem, an in-browser system to automatically
detect and block malicious third-party content inclusions before they can attack
the user's browser. Our system is complementary to other defensive approaches
such as CSP and Google Safe Browsing, and is implemented as a set of
modifications to the Chromium browser. \thesystem does not perform any
blacklisting to detect malicious third-party inclusions. Rather, it
incrementally constructs an inclusion tree for a given web page and
automatically prevents loading malicious resources by classifying their
inclusion sequences using a set of pre-built models.

Our evaluation over an 11 month crawl of the Alexa Top 200K demonstrates that
the prototype implementation of \thesystem detects a significant number of
malicious third-party content in the wild. In particular, the system achieved a
93.39\% detection rate with a false positive rate of 0.59\%. \thesystem was also
able to detect previously unknown malicious inclusions. We also evaluated the
performance and usability of \thesystem when browsing popular websites, and show
that the approach is capable of improving the security of users on the Web by
detecting 31 malicious inclusions during a user study without significantly
degrading the user experience.

\section*{Acknowledgement}

This material is based upon work supported by the National Science Foundation
under Grant No. CNS-1409738.

\bibliography{paper}{}
\bibliographystyle{plain}

\end{document}